\documentclass[fleqn,twoside]{article}

\usepackage{graphicx}
\usepackage[figuresright]{rotating}

\newcommand{\AmS}{{\protect\the\textfont2
  A\kern-.1667em\lower.5ex\hbox{M}\kern-.125emS}}
\usepackage{epsf}
\usepackage{epsfig}
\usepackage{soul}
\usepackage{amsmath}
\usepackage{amssymb}
\usepackage{amsbsy}
\usepackage{amsfonts}
\usepackage{graphics}
\usepackage{latexsym}
\usepackage{cite}
\usepackage[mathscr]{eucal}

\usepackage{espcrc2}

\newcommand{\bc}{\begin{center}}
\newcommand{\ec}{\end{center}}
\newcommand{\be}{\begin{equation}}
\newcommand{\ee}{\end{equation}}
\newcommand{\bea}{\begin{eqnarray}}
\newcommand{\eea}{\end{eqnarray}}

\newcommand{\eps}{\epsilon}
\newcommand{\ts}{\tau}

\def\Nf{n_f}

\newcommand{\nslash}{\not{\hspace{-0.08cm}n}}

\newcommand{\Dlr}{\overset{\leftrightarrow}{D}}

\title{Nucleon structure from generalized parton distributions
in lattice QCD\thanks{Talk presented by Ph. H\"agler.}\thanks{Preprint Edinburgh 2005/21, DESY 05-245}}

\author{M. G\"ockeler\address[reg]{Institut f\"ur Theoretische Physik,
Universit\"at Regensburg, 93040 Regensburg, Germany},
 Ph.~H\"agler\address[mun]{Institut f\"ur Theoretische Physik T39,
Physik-Department der TU M\"unchen,
James-Franck-Stra\ss{}e, \\
D-85747 Garching, Germany},
R.~Horsley\address[ed]{School of Physics,
University of Edinburgh,
Edinburgh EH9 3JZ, UK}, 
D.~Pleiter\address[nic]{John von Neumann-Institut f\"ur Computing NIC / DESY,
15738 Zeuthen, Germany},
P.E.L. Rakow\address[liv]{Theoretical Physics Division,
Department of Mathematical Sciences,
University of Liverpool,\\
Liverpool L69 3BX, UK},
A. Sch\"afer\addressmark[reg],
G.~Schierholz\addressmark[nic]\address[des]{Deutsches Elektron-Synchrotron DESY,
22603 Hamburg, Germany}
and J.M. Zanotti\addressmark[ed]\\QCDSF-UKQCD Collaborations}

\begin{document}

\begin{abstract}
This talk presents results from the QCDSF-UKQCD collaboration for
moments of leading twist generalized parton distributions in two-flavor
lattice QCD based on $\mathcal{O}(a)$ improved Wilson Fermions.
We study helicity independent and helicity flip
GPDs with a focus on densities of quarks in the transverse plane.
\vspace{-2mm}
\end{abstract}

\maketitle

\section{INTRODUCTION}
Generalized parton distributions (GPDs) \cite{GPD,Diehl:2003ny} enable us
to study many fundamental aspects of the intrinsic hadron structure
in the framework of QCD. They seamlessly link the at first sight completely
different concepts of parton distributions and form factors.
Furthermore, they allow us to define and calculate orbital
angular momentum contributions of quarks and gluons to the nucleon spin \cite{Ji:1996ek}.
Particularly important for this study is the interpretation of GPDs as probability densities
in impact parameter space \cite{Burkardt:2000za}.
The GPDs $H$ and $E$ of quarks in the nucleon are defined
by the following off-forward nucleon matrix elements \vspace{-1mm}
\begin{gather}
 \left\langle P',\Lambda ' \right|
  \!\int \!\!\frac{d \lambda}{4 \pi} e^{i \lambda x}
  \bar q \left(\!-\frac{\lambda}{2}n\!\right)
  \nslash
  {\cal U} q\left(\!\frac{\lambda}{2} n\!\right)
  \left| P,\Lambda \right \rangle
  = \; \nonumber\\
\,\,  \overline u(P',\Lambda ') \bigg( \!\!\!\nslash  H(x, \xi, t)  \nonumber\\
 \,\,\,\, + \frac{i \sigma^{\mu \nu} n_\mu \Delta_\nu} {2 m} E(x, \xi, t) \!\bigg)   u(P,\Lambda)\,,
\label{vectorGPDs}
\end{gather}
and the tensor GPDs are defined by \cite{Diehl:2001pm,Diehl:2005ev}
\begin{gather}
\left\langle
   P',\Lambda ' \right|\!\!\int \!\!\frac{d \lambda}{4 \pi} e^{i \lambda x}
  \bar q\!\left(\!\!-\frac{\lambda}{2}n\!\right)
  \, \!\!n_\mu\sigma^{\mu\nu}\gamma_5
  \mathcal{U}q\!\left(\!\frac{\lambda}{2} n\!\right)  \!\!
  \left| P,\Lambda \right\rangle = \nonumber\\
  \overline u(P',\Lambda ') n_\mu\bigg\{  \!
   \sigma^{\mu\nu}\gamma_5
   \bigg( \!\!H_T(x, \xi, t) - \frac{t}{2m^2}\widetilde H_T(x, \xi, t) \!\bigg)\nonumber\\
\,\,\,\, + \frac{\eps^{\mu\nu\alpha\beta} \Delta_{\alpha} \gamma_{\beta}} {2 m} \overline E_T(x, \xi, t) \nonumber\\
\,\,\,\, + \frac{\Delta^{[\mu} \sigma^{\nu]\alpha}\gamma_5 \Delta_{\alpha}} {2m^2 } \widetilde H_T(x, \xi, t)\nonumber\\
\,\,\,\, + \frac{\eps^{\mu\nu\alpha\beta} \overline P_{\alpha} \gamma_{\beta}} {m} \widetilde E_T(x, \xi, t)\!\bigg\}
 u(P,\Lambda)
 \label{GPDs1}\, ,
\end{gather}
where $f^{[\mu\nu]}=f^{\mu\nu}-f^{\nu\mu}$, $\Delta=P'-P$ is the momentum transfer with
$t=\Delta^2$, $\overline P = (P'+P)/2$, and $\xi=-n\cdot \Delta/2$
defines the longitudinal momentum transfer with the light-like vector $n$.
The Wilson line ensuring gauge invariance of the bilocal operator is denoted by $\mathcal{U}$.
Here and in the following we do not show explicitly the dependence of the GPDs on the
resolution scale $Q^2$.



A prominent feature of GPDs is that they reproduce the well-known
parton distributions in the forward limit, $\Delta=0$.
The relations to the unpolarized $q(x)$ and transversity $\delta q(x)$
parton distributions are given by $H(x,0,0)=q(x)=f_1(x)$ and $H_T(x,0,0)=\delta q(x)=h_1(x)$,
respectively.
%
On the other hand, integrating the GPDs over $x$ gives the form factors:
\begin{gather}
\int_{-1}^{1} dx H(x, \xi, t) = F_1(t)\, , \nonumber\\
\int_{-1}^{1} dx H_T(x, \xi, t) = g_T(t)\,.
\end{gather}
Another property of GPDs important for our investigations below
is their interpretation as densities in the transverse plane for $\xi=0$ \cite{Burkardt:2000za}.
To give an example, it has been shown that the impact parameter
dependent quark distribution for the quark GPD $H_q$,
\begin{gather}
q(x,b_\perp)
\equiv\int \frac{d^2\Delta_\perp}{(2\pi)^2} e^{-i b_\perp \cdot \Delta_\perp} H_q(x,0,t=-\Delta_\perp^2),
\label{impact}
\end{gather}
has the interpretation of a probability density for unpolarized quarks of flavor $q$ with longitudinal momentum fraction
$x$ and transverse position $b_\perp=(b_x,b_y)$ relative to the center of momentum in a nucleon.

\section{LATTICE CALCULATION OF\\GENERALIZED FORM FACTORS}
In order to facilitate the computation of the GPDs in lattice QCD,
we first transform the LHS of Eq.~(\ref{vectorGPDs},\ref{GPDs1}) to Mellin
space by forming the integral $\int_{-1}^{1} dx x^{n-1}\cdots$.
This gives off-forward nucleon matrix elements of towers of local operators
\begin{gather}
{\mathcal O}^{\mu \mu_{1}\ldots \mu _{n-1}}= \bar{q}(0)
\gamma^{\{\mu} i\Dlr{}^{\mu_{1}}\ldots
i\Dlr{}^{\mu_{n-1}\}}q(0)\, ,    \nonumber\\
{\mathcal O}_T^{\mu\nu \mu_{1}\ldots \mu _{n-1}}= \bar{q}(0)
\sigma^{\mu\{\nu}\gamma_5 i\Dlr{}^{\mu_{1}}\ldots
i\Dlr{}^{\mu_{n-1}\}}q(0)\, ,
\label{eq:ops}
\end{gather}
where $\Dlr = \frac{1}{2}(\overrightarrow{D} -
\overleftarrow{D})$ and $\{\cdots\}$ indicates symmetrization of
indices and subtraction of traces.
Matrix elements of the operators in Eq.(\ref{eq:ops}) are parametrized in terms of generalized
form factors (GFFs)  $A_{ni}$, $B_{ni}$ and $C_{n0}$ for the vector, and
$A_{Tni}$, $\overline B_{Tni}$, $\widetilde A_{Tni}$ and
$\widetilde B_{Tni}$ for the tensor case.
%
For the vector operators, the parametrizations for the lowest three moments read
\begin{gather}
\left\langle P^{\prime}\Lambda^{\prime}\right|
 \bar{q}(0) \gamma^{\mu} q(0)
\left| P \Lambda\right\rangle
=\nonumber\\
\,\,\overline u(P',\Lambda ') \bigg\{ \gamma^{\mu}A_{10}(t)
+\frac{i \sigma^{\mu \nu} \Delta_\nu} {2 m} B_{10}(t) \bigg\} u(P,\Lambda)\ , \label{paraA1} \displaybreak[0]\\
\vspace{5mm}
\left\langle P^{\prime}\Lambda^{\prime}\right|
 \bar{q}(0) \gamma^{\{\mu} i\Dlr{}^{\mu_{1}\}}q(0)
\left| P \Lambda\right\rangle
=\nonumber\\
\,\,\overline u(P',\Lambda ') \bigg\{ \overline P^{\{\mu}\gamma^{\mu_{1}\}}A_{20}(t)
+\overline P^{\{\mu}\frac{i \sigma^{\mu_{1}\}\nu} \Delta_\nu} {2 m} B_{20}(t) \nonumber\\
\,\,\,\,+\frac{1}{m}\Delta^{\{\mu} \Delta^{\mu_{1}\}} C_{20}(t)\bigg\} u(P,\Lambda)\ ,\displaybreak[0]\\
\intertext{and}
\vspace{5mm}
\left\langle P^{\prime}\Lambda^{\prime}\right|
 \bar{q}(0) \gamma^{\{\mu} i\Dlr{}^{\mu_{1}} i\Dlr{}^{\mu_{2}\}}q(0)
\left| P \Lambda\right\rangle
=\nonumber\\
\,\,\overline u(P',\Lambda ') \bigg\{ \overline P^{\{\mu}\overline P^{\mu_1}\gamma^{\mu_{2}\}}A_{30}(t) \nonumber\\
\,\,\,\,+\overline P^{\{\mu}\overline P^{\mu_1}\frac{i \sigma^{\mu_{2}\}\nu} \Delta_\nu} {2 m} B_{30}(t)
+\Delta^{\{\mu}\Delta^{\mu_1}\gamma^{\mu_{2}\}}A_{32}(t) \nonumber\\
\,\,\,\,+\Delta^{\{\mu} \Delta^{\mu_1}\frac{i \sigma^{\mu_{2}\}\nu} \Delta_\nu} {2 m} B_{32}(t) \bigg\} u(P,\Lambda)\ .
\label{para1}
\end{gather}
%
For the $n=1$ helicity flip operator, we have \cite{Diehl:2001pm,Hagler:2004yt}
\begin{gather}
\left\langle P^{\prime}\Lambda^{\prime}\right| 
 \bar{q}(0) \sigma^{\mu\nu}\gamma_5 q(0)
\left| P \Lambda\right\rangle
=\nonumber\\
\,\,\overline u(P',\Lambda ') \bigg\{ \sigma^{\mu\nu}\gamma_5
   \bigg( A_{T10}(t) - \frac{t}{2m^2}\widetilde A_{T10}(t) \bigg)\nonumber\\
\,\, \,\, + \frac{\eps^{\mu\nu\alpha\beta} \Delta_{\alpha} \gamma_{\beta}} {2 m} \overline B_{T10}(t)\nonumber\\
\,\,\,\, +  \frac{\Delta^{[\mu} \sigma^{\nu]\alpha}\gamma_5 \Delta_{\alpha}} {2m^2 } \widetilde A_{T10}(t)\!\bigg\}
 u(P,\Lambda)\ .
\label{tn0}
\end{gather}
Parametrizations for higher moments $n\ge 1$ can be found in \cite{Hagler:2004yt,Chen:2004cg}.
%
%

Lattice counterparts of the continuum Minkowski space-time matrix elements
in Eqs.~(\ref{paraA1}-\ref{tn0}) are nucleon two- and three-point
correlation functions within
a discretized Euclidean space-time framework.
%
They are defined by
\begin{gather}
  C^{\text{2pt}}(\tau,P) = \sum_{j,k}
    \left(\!\frac{1\!+\!\gamma_4}{2}\!\right)_{\!\!jk}\!\!\left\langle N_{k}
    (\tau,P) \overline{N}_{j}(\tau_{\text{src}},P)\right\rangle,
    \displaybreak[0]\\
\intertext{and}
  C_{\cal O}^{\text{3pt} \mu_{1}\ldots
    \mu_{n-1}}(\tau,P',P) = \nonumber\\
 \,\,   \sum_{j,k}
    \tilde\Gamma_{jk}\left\langle N_{k}
    (\tau_{\text{snk}},P'){\cal O}^{ \mu_{1}\ldots \mu _{n-1}}(\tau)
  \overline{N}_{j}(\tau_{\text{src}},P)\right\rangle,
  \label{threept}
\end{gather}
where $\tilde \Gamma$ is a (spin-)projection matrix and the operators
$\overline{N}$ and $N$ create and destroy states with the quantum
numbers of the nucleon, respectively.
%
%
Inserting complete sets of states and using the time
evolution operator, we can rewrite Eq.~(\ref{threept}),
\begin{gather}
C_{\cal O}^{\text{3pt} \mu_{1}\ldots
    \mu_{n-1}}(\ts,P',P) =\nonumber\\
 \,\, \frac{\left(Z(P)\overline{Z}(P')\right)^{1/2}}{4 E(P')E(P)}
  e^{-E(P)\ts-E(P')(\ts_{\text{snk}}-\ts)}
   \nonumber\\
  \,\,\,\,  \times \sum_{\Lambda,\Lambda^{\prime }}
    \left\langle P',\Lambda^{\prime }\right|{\cal O}^{\mu_{1}\ldots \mu _{n-1}}
    \left|P,\Lambda \right\rangle \nonumber\\
\,\,\,\,    \times\overline{u}(P,\Lambda)\tilde\Gamma
    u(P',\Lambda')+\ldots\ ,
 \label{threept2}
\end{gather}
where we set $\ts_{\text{src}}=0$.
The two-point function can be written as
\begin{gather}
C^{\text{2pt}}(\tau,P)=\nonumber\\
\,\,\left(Z(P)\overline{Z}(P)\right)^{1/2}
 \frac{E(P)+m}{E(P)}
 e^{-E(P)\ts} + \ldots\ .
\label{twopt2}
\end{gather}
Here and below, $m$ denotes the mass of the nucleon ground state.
Excited states with energies $E'>E(P),E(P')$ are represented by the ellipsis in Eq.~(\ref{threept2}) and (\ref{twopt2}).
They are exponentially suppressed as long as $\ts\gg 1/E', \ts_{\text{snk}}-\ts\gg 1/E'$.
%
Using the explicit parametrizations from Eqs.~(\ref{paraA1}-\ref{tn0})
transformed to Euclidean space together with Eq.~(\ref{threept2}), we can sum over polarizations
to obtain
\begin{gather}
  \label{threept3}
 C_{\cal O}^{\text{3pt}\mu_{1}\ldots
    \mu_{n-1}}(\ts,P',P)
  =\nonumber\\
 \,\, \frac{\left(
      Z(P)\overline{Z}(P')\right)^{1/2}}{4 E(P')E(P)}
 e^{-E(P)\ts-E(P')(\ts_{\text{snk}}-\ts)}\nonumber\\
\,\,\,\,   \times\text{Tr}\bigg[ \tilde\Gamma(i\!\!\!\not{\!P}^{\prime}-m)
   \bigg(a^{ \mu_{1}\ldots \mu _{n-1}}A_{n0}(t)\nonumber\\
\,\,\,\, +   b^{ \mu_{1}\ldots \mu _{n-1}}B_{n0}(t)+\cdots \bigg)
   (i\!\!\!\not{\!P}-m)
  \bigg]\ ,
  \label{trace}
\end{gather}
\begin{figure}[t]
\bc
\includegraphics[height=6.5cm,angle=-90]{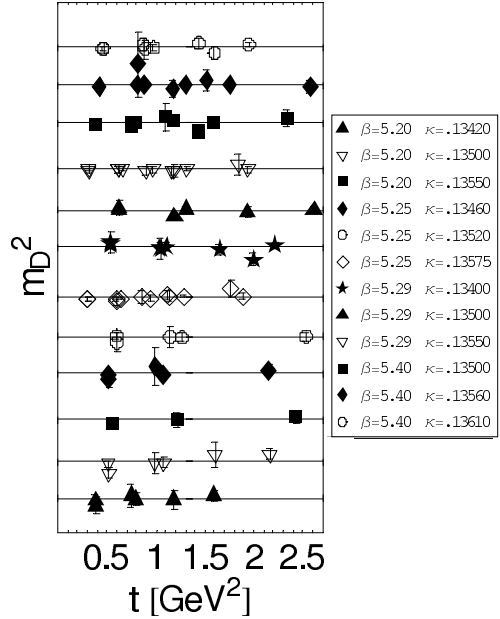}
\caption{Effective dipole masses for the iso-vector form factor $A_{10}=F_1$ for all datasets.}
\label{effdipole}
\ec
\vspace{-.5cm}
\end{figure}
where $a,b\ldots$ are the Euclidean versions of the corresponding
prefactors in Eqs.~(\ref{paraA1}-\ref{tn0}).
The Dirac-trace in Eq.~(\ref{trace}) is evaluated explicitly, while
the normalization factor and the exponentials
are cancelled by constructing a
ratio $R$ of two- and three-point functions,
\begin{gather}
R_{\cal O}(\tau,P', P)\, =
 \frac{C^{\text{3pt}}_{\cal O} (\tau,P', P)}
        {C^{\text{2pt}}(\tau_{\text{snk}},P')}\nonumber\\
\times \left[
  \frac{C^{\text{2pt}}(\tau,P') C^{\text{2pt}}(\tau_{\text{snk}},P')
     C^{\text{2pt}}(\tau_{\text{snk}}\!-\!\tau,P)}
       {C^{\text{2pt}}(\tau,P) C^{\text{2pt}}(\tau_{\text{snk}},P)
     C^{\text{2pt}}(\tau_{\text{snk}}\!-\!\tau,P')}
\right]^{\frac{1}{2} }.
\label{eq:ratio}
\end{gather}
The ratio $R$ is averaged over the plateau region and then equated with the
corresponding sum of GFFs times $P$- and $P'$-dependent calculable
pre-factors.
For a given moment $n$, this is done simultaneously for all
contributing index combinations $(\mu_{1}\ldots \mu _{n-1})$
and all discrete lattice momenta $P,P'$ corresponding to the same
value of $t=(P'-P)^2$.
Following this procedure, we find in general an overdetermined set of
equations from which we extract the GFFs \cite{QCDSF-1,MIT}.
On the lattice the space-time symmetry is reduced to the hypercubic
group $H(4)$, and the lattice operators have to be chosen such that
they belong to irreducible multiplets under $H(4)$. For our analysis of the
$n=1,2,3$-vector operators, we employ  all operators belonging to the
corresponding $4$-~, $9$- and $12$-dimensional spaces described in \cite{Gockeler:1996mu},
and for the $n=1$ tensor operators we utilize the 6-dimensional multiplet as discussed in \cite{Diehl:2005ev}.
The statistical error on the GFFs is obtained from a jackknife analysis.
Our results have been non-perturbatively renormalized \cite{reno}
and transformed to the $\overline{\mbox{MS}}$ scheme at a scale of $4$ GeV$^2$.
\begin{figure}[t]
\bc
\includegraphics[height=7.cm,angle=-90]{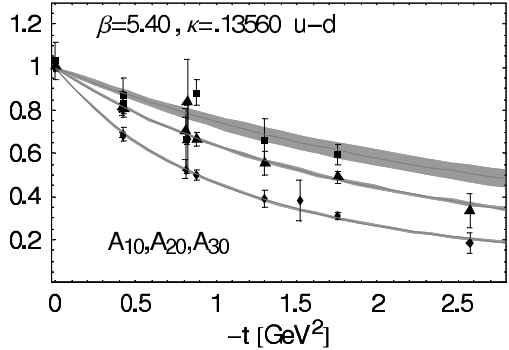}
\caption{Slopes of the generalized form factors $A_{10}(t)$ (lowest curve),
$A_{20}(t)$ and $A_{30}(t)$ (highest curve).}
\label{slopes}
\ec
\vspace{-1cm}
\end{figure}
The lattice results to be discussed below have been obtained from simulations with 
$\Nf=2$ flavors of dynamical non-perturbatively ${\mathcal O}(a)$ improved Wilson fermions and Wilson glue.
There are 12 datasets available consisting of four different couplings
$\beta=5.20$, $5.25$, $5.29$, $5.40$ with three different $\kappa=\kappa_{\mathrm {sea}}$ values per $\beta$.
The pion masses of our calculation vary from $550$ to $1000$ MeV, and
the lattice spacings and spatial volumes vary between 0.07-0.11~fm and (1.4-2.0~fm)$^3$ respectively.
Our simulations are based on three sink momenta,
$\vec{p}_0 = ( 0, 0, 0 )$,
$\vec{p}_1 = ( p, 0, 0 )$,
$\vec{p}_2  = ( 0, p, 0 )$
($p=2\pi/L_S$) and three projection matrices,
$\tilde\Gamma_{\rm unpol} = \frac{1}{2}(1+\gamma_4)$,
$\tilde\Gamma_1 = \frac{1}{2}(1+\gamma_4)\, i\gamma_5\gamma_1$,
$\tilde\Gamma_2 = \frac{1}{2}(1+\gamma_4)\, i\gamma_5\gamma_2$.
Our present calculation does not include the computationally demanding disconnected contributions. We expect,
however, that they are small for the tensor GFFs \cite{Gockeler:2005cj}.

%
\begin{figure}[t]
\bc
\includegraphics[height=7.cm,angle=-90]{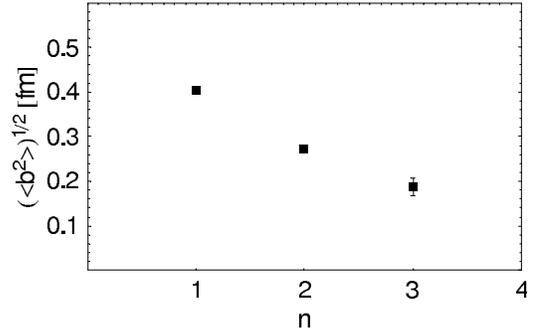}
\caption{Decrease of the charge radius with increasing $n$.}
\label{radius}
\ec
\vspace{-1cm}
\end{figure}
\section{LATTICE RESULTS FOR THE \\LOWEST MOMENTS OF QUARK \\DENSITIES IN THE NUCLEON}
In a finite volume, momenta are discretized and we obtain the GFFs only for a limited number
of different values of the momentum transfer squared $t$. In our simulation, there are
in general 16 $t$-values available per dataset in a range of $0\le t < 4$ GeV$^2$. 
In order to get a representation of the GFFs for continuous values of $t$,
we parametrize them using a $p$-pole ansatz
\begin{equation}
 F(t)=\frac{F(0)}
  {\left( 1 - {t/m_p^2}
   \right)^p} \ ,
\label{ppole}
\end{equation}
where the values of $F(0)$, $m_p$ and $p$ for the individual GFFs are fixed by a fit to the lattice results.
The ansatz in Eq.~(\ref{ppole}) is then Fourier transformed in order to get the GFFs in coordinate space.
For the vector GFFs $A_{10}$, $A_{20}$ and $A_{30}$, we use a dipole ansatz, $p=2$.
More details of the $p$-pole parametrization and some numerical results for the
parameters can be found in \cite{Gockeler:2004mn,Gockeler:2004vx,Gockeler:2005aw}.
Let us note that the values for the power $p$ we are
using for the different GFFs lead to a regular behavior of the quark densities
in the limit $b_\perp\to 0$, as discussed in \cite{Diehl:2005jf}. In order to check
our dipole ansatz, we invert Eq.(\ref{ppole}) and show in Fig.(\ref{effdipole}) the effective  dipole-mass squared
$m_{D}^2(t)=m_{p=2}^2(t)$ as a function of the momentum transfer squared $t$ for the $A_{10}(t)=F_1(t)$ form factor.
The horizontal lines indicate the results for $m_{D}^2$ from fits  to the individual datasets using Eq.~(\ref{ppole}).
The plot shows that the dipole represents an accurate description of the form factor for a large
range of the momentum transfer.
\begin{figure*}[t]
\bc
\includegraphics[height=15cm,angle=-90]{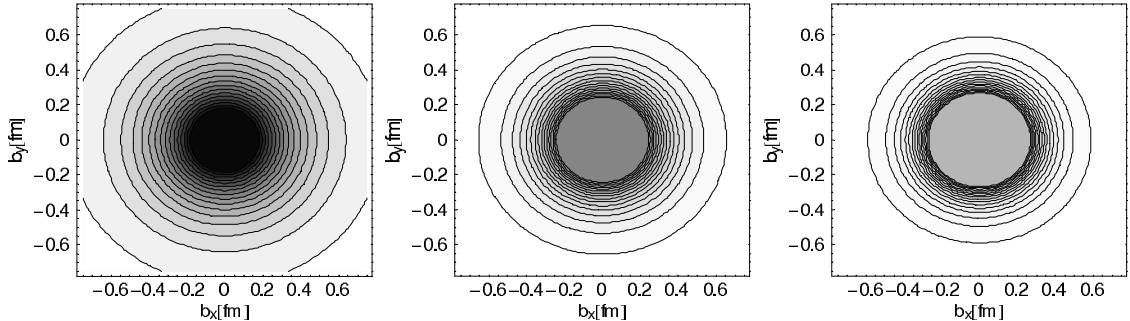}
\caption{Spin independent densities of up-quarks in the nucleon for $n=1,2,3$ (from left to right).}
\label{densities1}
\ec
\vspace{-.5cm}
\end{figure*}
%

A momentum transfer squared of the order of a few GeV$^2$ or more
has in general to be redistributed between the partons in order to prevent a break-up of
the nucleon. According to the Feynman mechanism, parton configurations for which the active
quark has a large longitudinal momentum fraction may still form a bound state after the scattering
without a redistribution of the momentum transfer. This simple picture of the scattering mechanism
is confirmed by our results in Fig.(\ref{slopes}), where we compare the slope in $t$
of the generalized form factors $A_{n0}$ for the lowest three moments, 
which have been normalized to unity at $t=0$ GeV$^2$.
To see this most clearly, let us first note that higher moments $n$ correspond to larger average
momentum fractions $\overline x^n \approx \langle x^n\rangle/\langle x^{n-1}\rangle$ \cite{Renner:2005sm}.
Fig.(\ref{slopes}) shows that already at a momentum transfer squared of $t\approx -2$ GeV$^2$
the (normalized) form factor $A_{30}$ is a factor of two larger than the form factor $A_{10}$.
This indicates that a momentum transfer of this magnitude to the nucleon can be better
absorbed by partons with large momentum fraction $x$, thereby lowering the probability
of a break-up of the bound state. In the extreme case $n\to \infty$, we expect the generalized
form factors to be independent of $t$, $A_{n\to \infty, 0}(t)/A_{n\to \infty, 0}(0)=1$.
The flattening of the slope for the $A_{n0}$ GFFs in lattice QCD
has for the first time been observed in \cite{MIT-2}.

\begin{figure*}[t]
\bc
\includegraphics[height=15cm,angle=-90]{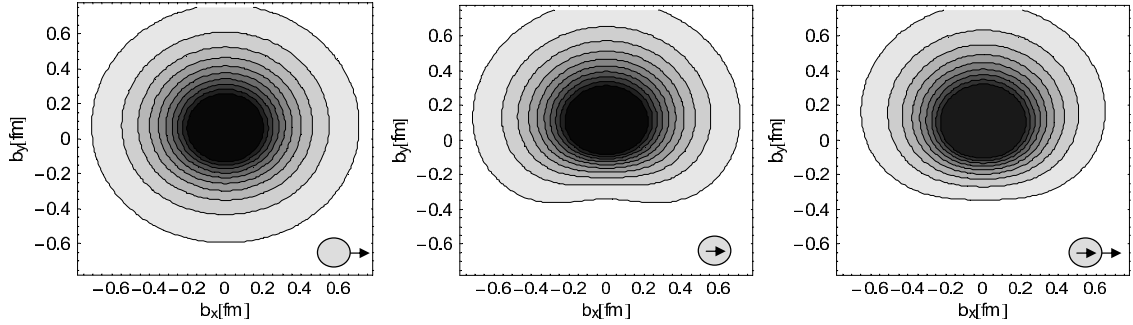}
\caption{Densities of up-quarks in the nucleon.
The nucleon and quark spins are oriented in the transverse plane as indicated, where
the inner arrow represent the quark and the outer arrow the nucleon spin.
A missing arrow represents the unpolarized case.}
\label{densities2}
\ec
\end{figure*}
In impact parameter space, the increase of the dipole masses translates into a
decrease of the charge radius for increasing average momentum fraction.
Fig.(\ref{radius}) shows that the charge radius of the nucleon in the transverse plane drops by a factor
of about two, going from $n=1$, which corresponds to an average momentum fraction of
$\overline x^{n=1}\approx 0.25$ to $n=3$, where $\overline x^{n=3}\approx 0.4$.
Since the impact parameter measures the distance of the active quark to the
center of momentum $R_\perp$, we expect the distribution of the quarks in the transverse
plane to peak around $R_\perp=0_\perp$ when $x\to 1$ \cite{Burkardt:2002hr}.

The $x$-moments of the impact parameter dependent quark distribution in Eq.(\ref{impact}) are given by the
GFFs in impact parameter space,
\begin{gather}
\int_{-1}^{1} x^{n-1}q(x,b_\perp) = A_{n0}(b_\perp)\,.
\label{impact2}
\end{gather}
Fourier-transforming the $p$-pole ansatz for the GFFs in Eq.(\ref{ppole}) to impact parameter space,
we can calculate the distribution of quarks in the transverse plane for the lowest three
moments we consider in this work. The result is shown in Fig.(\ref{densities1}), where we plot
densities of up-quarks in Eq.(\ref{impact2}) for $n=1,2,3$.
To be precise, let us note that the plots do not exactly show probability densities because the moments
$\int_{-1}^{1}x^{n-1}$ always correspond to differences/sums of quark and anti-quark densities.
The quark distributions in Fig.(\ref{densities1}) clearly get narrower from the left ($n=1$) to the right ($n=3$) plot.

In the following, we will discuss the recently observed correlation of transverse quark spin and
impact parameter which shows up in transverse spin densities of quarks in the nucleon \cite{Diehl:2005jf}.
It turns out that these correlations in the transverse plane
are governed by quark helicity flip (or tensor) GPDs.
The lowest moment of the quark transverse spin density reads \cite{Diehl:2005jf}
\begin{gather}
\left\langle P^+,R_\perp=0,S_\perp \right|
\hat\rho_T^{n=1}(b_\perp)
\left| P^+,R_\perp=0,S_\perp \right\rangle
= \nonumber\\
\,\,\frac{1}{2}\left\{
  A_{10}(b_\perp)
+ s_\perp^i S_\perp^i \bigg( A_{T10}(b_\perp)\right.  \nonumber\\
\,\,\,\,- \frac{1}{4m^2} \Delta_{b_\perp} \widetilde{A}_{T10}(b_\perp) \bigg)\nonumber\\
\,\,\,\,+ \frac{ b_\perp^j \eps ^{ji}}{m} \left( S_\perp^i B_{10}'(b_\perp)
+ s_\perp^i \overline{B}_{T10}'(b_\perp) \right) \nonumber\\
\,\,\,\,+ \left. s_\perp^i ( 2 b_\perp^i b_\perp^j
- b_\perp^2 \delta^{ij} ) S_\perp^j \frac{1}{m^2} \widetilde{A}_{T10}''(b_\perp)
\right\},
  \label{density}
\end{gather}
where the lowest moment of the transverse spin density operator is given by
\begin{eqnarray}
\hat\rho_T^{n=1}(b_\perp)
= \frac{1}{2} \overline q(b_\perp)\big[\gamma^+
- s_\perp^j i \sigma^{+j}\gamma_5\big]q(b_\perp)
  \label{densityOp}
\end{eqnarray}
which projects out quarks with transverse polarization $s_\perp = (\cos\chi',\sin\chi')$.
The transversity states
\begin{gather}
\left| P^+,R_\perp=0,S_\perp \right\rangle =
\frac{1}
{\sqrt{2}}\Big(\left| P^+,R_\perp=0,\Lambda=+\right\rangle \nonumber\\
\,\,\,\,+ e^{i\chi}\left| P^+,R_\perp=0,\Lambda=-\right\rangle\Big)
\end{gather}
describe a nucleon with longitudinal momentum $P^+=(P^0+P^3)/\sqrt{2}$ which is localized in the transverse
plane at $R_\perp=0$ and has transverse spin $S_\perp = (\cos\chi,\sin\chi)$.
The impact parameter dependent GFFs in Eq.~(\ref{density}) are the Fourier-transforms
of the momentum space GFFs at $\xi=0$, as in Eq.~(\ref{impact}).
The derivatives in Eq.~(\ref{density}) are defined by $f'(b_\perp) \equiv \partial_{b_\perp^2}f(b_\perp)$ and
$\Delta_{b_\perp}f(b_\perp) \equiv 4\partial_{b_\perp^2}\big(b_\perp^2\partial_{b_\perp^2}\big)f(b_\perp)$.
%
%
In Fig.(\ref{densities2}) we show preliminary results
for the lowest moment of transverse spin densities of quarks in the nucleon for up quarks.
The densities are strictly positive for all $b_\perp$, indicating that the contributions from anti-quarks are small.
On the LHS of Fig.(\ref{densities2}), we show the density of unpolarized quarks in a transversely polarized nucleon.
The distortion of the density in $+y$-direction coming from the term $\propto B_{10}'(b_\perp)$
has already been discussed in \cite{Burkardt:2002hr}. In \cite{lensing}, a formalism has been developed 
which predicts that such a distortion leads to single spin asymmetries and in particular to a
non-zero negative Sivers function \cite{Sivers:1989cc} $f_{1T}^\perp<0$ for up-quarks. 
First results from experiment seem
to confirm this \cite{Airapetian:2004tw}.
The density plot in the center of Fig.(\ref{densities2}) shows that the term
$\propto \overline B_{T10}'(b_\perp)$ in Eq.(\ref{density})
also leads to a strong transverse deformation orthogonal to the transverse quark
spin for an \emph{unpolarized} nucleon.
According to \cite{Burkardt:2005hp}, this distortion in
$+y$-direction\footnote{or equivalently a shift in $(-x)$-direction for quarks with spin in $y$-direction}
due to the transversely polarized quarks may correspond to a non-zero, 
negative Boer-Mulders function \cite{Boer:1997nt} $h_1^{\perp}<0$ for up-quarks.
For the plot on the RHS in Fig.(\ref{densities2}), the distortions due to transverse quark
and nucleon spin add up,
while there is practically no influence visible from the quadrupole-term $\propto s_\perp^i ( 2 b_\perp^i b_\perp^j
- b_\perp^2 \delta^{ij} ) S_\perp^j$ in Eq.~(\ref{density}).
%
%
%
\section{CONCLUSIONS AND OUTLOOK}

We presented results for the lowest moments of helicity independent and 
helicity flip generalized parton distributions in $\Nf=2$ lattice QCD.
Already at a momentum transfer of a few GeV$^2$, we find that the vector coupling
to the nucleon is noticeably stronger for quarks with larger momentum fraction.
Equivalently, the charge radius of the nucleon decreases visibly 
with increasing average momentum fraction.
For transverse polarizations, we observe strong distortions
of the spin densities orthogonal to the direction of the quark and nucleon spin.
These distortions could give rise to non-vanishing Sivers and Boer-Mulders functions
through final state interactions as argued by Burkardt \cite{Burkardt:2005hp,lensing}.

We plan to extend the analysis of moments of generalized parton
distributions, including a study of the lowest two moments of transverse spin densities
for up- and down-quarks, and to investigate improved positivity bounds for GPDs which have
been obtained in \cite{Diehl:2005jf}.

\section*{Acknowledgments}

The numerical calculations have been performed on the Hitachi SR8000
at LRZ (Munich), on the Cray T3E at EPCC (Edinburgh) 
\cite{Allton:2001sk}, and on the APEmille at NIC/DESY
(Zeuthen). This work is supported by the DFG (Forschergruppe
Gitter-Hadronen-Ph\"anomenologie and Emmy-Noether-program), 
by the EU I3HP under contract number RII3-CT-2004-506078 and by the
Helmholtz Association, contract number VH-NG-004.

\end{document}